\documentclass[12pt]{article}
\usepackage[T1]{fontenc} 
\usepackage[utf8x]{inputenc}
\usepackage{enumerate}
\usepackage{lastpage}
\usepackage{fancyvrb}
\usepackage{pdfpages} 
\usepackage[a4paper,margin=3.2cm]{geometry}
\usepackage{fancyhdr}
\usepackage{calc} 
\usepackage{caption}
\usepackage{tabularx}
\usepackage{newfloat} 
\usepackage{morefloats}  
\usepackage{subcaption}
\usepackage{hyperref}
\usepackage{amsmath, amsthm}
\usepackage{amsfonts}
\usepackage{amssymb}
\usepackage{graphicx}
\usepackage{textcomp}
\usepackage{hyperref}
\usepackage{color, colortbl}
\usepackage{array}
\usepackage{enumitem}
\usepackage{multirow, makecell}
\usepackage{wrapfig}
\usepackage{multicol}
\usepackage{numprint}
\usepackage{algorithm, algpseudocode}
\usepackage{./customized}

\title{Automatic Repair of Infinite Loops}
\author{
\begin{minipage}{7cm}
\begin{center}
Sebastian R. Lamelas Marcote\\
University of Buenos Aires\\ Argentina 
\end{center}
\end{minipage}\hfill\begin{minipage}{7cm}
\begin{center}
Martin Monperrus\\
University of Lille \& INRIA\\ France
\end{center}
\end{minipage}}
\date{}

\begin{document} 

\maketitle

\begin{abstract}
Research on automatic software repair is concerned with the development of systems that automatically detect and repair bugs.  
One well-known class of bugs is the infinite loop. Every computer programmer or user has, at least once, experienced this type of bug. We state the problem of repairing infinite loops in the context of test-suite based software repair: given a test suite with at least one failing test, generate a patch that makes all test cases pass. Consequently, repairing infinites loop means having at least one test case that hangs by triggering the infinite loop. Our system to automatically repair infinite loops is called $Infinitel$. We develop a technique to manipulate loops so that one can dynamically analyze the number of iterations of loops; decide to interrupt the loop execution; and dynamically examine the state of the loop on a per-iteration basis. Then,  in order to synthesize a new loop condition, we encode this set of program states as a code synthesis problem using a technique based on Satisfiability Modulo Theory (SMT). We evaluate our technique  on seven seeded-bugs and on seven real-bugs. $Infinitel$ is able to repair all of them, within seconds up to one hour on a standard laptop configuration.
\end{abstract}

\section{Introduction}

Research on automatic software repair is concerned with the development of systems that automatically detect and repair bugs. 
We consider as bug a behavior observed during program execution that does not correspond to the expected one.
Automatic software repair is close to other research areas such as automatic debugging, software testing, program synthesis, machine learning for software engineering. 
There have been a number of results in this field (e.g. \cite{Arcuri2008,genprog,par,semfix}), since seminal work  at the end of the 2000ies \cite{Arcuri2008,genprog,dallmeier2009generating}.

The ultimate goal of automatic software repair is to minimize the maintenance costs.
Software maintenance is often considered the most expensive development phase \cite{trends} and a key task during maintenance is the correction of bugs (colloquially ``bug fixing''). 
The automatic repair of even a fraction of software bugs would translate to significant savings in developer time and costs.

Hamill and Goseva-Popstojanova \cite{trends} showed that one of the most common types of software faults are \it{coding faults}. That is, faults directly in the source code, according to a given set of requirements. For instance, incorrectly assigned values, uninitialized values, missing data validation, incorrect loop statements, and so on. 
We have recently argued that for devising sound repair techniques, we need a systematic taxonomy of common coding faults, so that we can develop an effective repair method for each type \cite{essay}. 
That is, to each bug corresponds a \it{defect class} and, to repair it, a specific repair method which exploits the defect class' intrinsic properties is used.

One well-known defect class is the \qt{infinite loop}. Every computer programmer or user has, at least once, experienced this type of bug. 
It's so much part of the programming folklore that Apple Inc. has renamed the street encircling its head quarters ``Infinite Loop''\footnote{see \url{https://www.apple.com/about/}, last visited April 19, 2015.}.
Infinite loops are  responsible for hanged  programs and frozen user interfaces. Technically, It consists of a loop which unintentionally iterates forever without returning an expected result or throwing an exception. In this paper, we aim to automatically repair this defect class.

We believe the infinite loop defect class is fairly common. Take, for instance, one of the historically most popular UNIX commands: \vb{grep}. \refex{grepCommit} shows the excerpt of a commit in \vb{grep}'s codebase\footnote{\url{http://git.savannah.gnu.org/cgit/grep.git/}}. As indicated by the commit message, the changes of the commit are made to fix an infinite loop. In order to do so, the boolean condition of the loop is corrected and a \vb{break} statement is introduced.

\bcenter
\begin{example}
\begin{batim}
# Commit message: Fix hang on 'grep --color "" anything'
\inred{- while ((match_offset = (*execute) (beg, lim - beg, &match_size, 1)) != (size_t) -1) \{}
\indarkgreen{+ while (lim-beg && (match_offset = (*execute) (beg,lim - beg,&match_size,1)) != (size_t) -1) \{}
  char const *b = beg + match_offset;
  /* Avoid matching the empty line at the end of the buffer. */
  if (b == lim)
    break;
  \indarkgreen{+ /* Avoid hanging on grep --color "" foo */}
  \indarkgreen{+ if (match_size == 0)}
    \indarkgreen{+ break;}
  fwrite (beg, sizeof (char), match_offset, stdout);
\end{batim}
\captionexample{Infinite loop patch in \vb{grep.c} (commit 3ec7191f).}{grepCommit}
\end{example}
\ecenter

In this paper, we propose a technique to automatically repair infinite loops.
To our knowledge, there is no published work on this topic.
We state the problem of repairing infinite loops in the context of test-suite based software repair \cite{essay}: given a test suite with at least one failing test, generate a patch that makes all test cases pass. 
Consequently, repairing infinites loop means having at least one test case that hangs by triggering the infinite loop.
However, the loop that is running infinitely may be executed a finite number of times in other test cases.
Hence, repairing an infinite loop means modifying the behavior of the infinite loop so that \emph{every test case} invoking the infinite loop both halts and passes. 
In our case, the patch we aim to synthesize consists of a new boolean expression for the loop condition of the infinite loop. In other words, for the repair to be successful, the new predicate must correct every infinite execution happening in the non-halting test cases and must also keep the already passing test cases correct.

Our system to automatically repair infinite loops is called $Infinitel$.
It is based on a technique we develop to manipulate loops so that one can 
dynamically analyze the number of iterations of loops;
decide to interrupt the loop execution;
and dynamically examine the state of the loop on a per-iteration basis. 
Then,  in order to synthesize a new loop condition, we encode this set of program states as a code synthesis problem using a technique based on Satisfiability Modulo Theory (SMT).

We evaluate $Infinitel$  on seven seeded-bugs and on seven real-bugs.
Our technique is able to repair all of them, within seconds up to one hour on a standard laptop configuration.
We deeply discuss those cases to understand the strength and weaknesses of our automatic repair technique.

To sum up, the contributions of this paper are:

\bi
\item The problem statement of automatic repair for infinite loops.
\item A source code instrumentation  technique to dynamically analyze the behavior of loops. 
\item An end-to-end repair algorithm for infinite loops, based on runtime loop state analysis and code synthesis.
\item The evaluation of the proposed solution with 7 seeded bugs and 7 real bugs.
\ei

The rest of the paper is organized as follows. In \refsection{loopTheory} we define a terminology for loops that is used throughout the paper. 
In \refsection{ourApproach}, we describe our solution to automatically repair the infinite loop defect class.
In \refsection{evaluation} we discuss the evaluation of our approach.
In \refsection{discussion}, we analyze the core assumptions behind our system.
In \refsection{related}, we compare our approach to other related work. We conclude the paper in \refsection{conclusions}.

\rsection{Loop Theory}{loopTheory}

\begin{example}
\centering
\begin{subexample}{.32\linewidth}
\begin{batim}




void clear(int[] array) \{
  int n = array.length;
  for (int i = 0; i < n; i++)
    array[i] = 0;
\}




\end{batim}
\captionexample{}{standard}
\end{subexample}
\begin{subexample}{.37\linewidth}
\begin{batim}
int index(int[] sorted, int e) \{
  int low = 0;
  int high = sorted.length - 1;
  do \{
    int mid = (low + high + 1) /2;
    if (sorted[mid] <= e) \{
      low = mid;
    \} else \{
      high = mid;
    \}
  \} while (sorted[low] != e);
  return low;
\}
\end{batim}
\captionexample{}{idempotent}
\end{subexample}
\begin{subexample}{.25\linewidth}
\begin{batim}
int method(int a) \{
  int b = a;
  while (b > 0) \{
    if (b == 18) \{
      return a;
    \}  
    if (b == 9) \{
      break;
    \}
    b -=1;
  \}
  return b;
\}
\end{batim}  
\captionexample{}{breakAndReturn}
\end{subexample}
\captionexample{
\subrefex{standard}: \vb{for} loop. 
\subrefex{idempotent} \it{idempotent} \vb{while} loop. \subrefex{breakAndReturn} \vb{while} loop with a \it{break} and \it{return} statements.}{fourLoopExamples}
\end{example}

In this section we introduce the concepts and terminology related to loops that will be used throughout the paper.  We also define the fault class we address.

\subsection{Terminology}
A \it{loop} is a control flow statement which permits to repeatedly execute a block of statements.
Classical loops are \vb{for}, \vb{while} or \vb{do-while} loops. 
The \it{loop body} refers to the block of statements inside the loop. 

The \it{looping guard} is the boolean condition used in a loop to control termination. For instance, the looping guard in \refex{standard} is \qvb{i < n}. The role of the looping guard is twofold. 
On one hand, before executing the first iteration of the loop, the looping guard acts as a \it{precondition} (but for do-while loops). If the precondition is not met (the first evaluation of the looping guard returns false), the flow of the program continues without entering the loop altogether. 
On the other hand, if the precondition is met, the first iteration begins and, from then on, the looping guard will be acting as an \it{exit condition}. If any subsequent evaluation of the looping guard evaluates to false, then the exit condition is met, and the flow can continue outside of the loop.

A loop may also terminate when certain instructions are executed within the loop body (break and return statements, exceptions) 
A \it{break statement} is an instruction that breaks the loop from within the loop body (e.g., statement of second \vb{if} in \refex{breakAndReturn}).
A \it{return statement} is an instruction that both breaks the loop and exits the function or method containing it (e.g., statement of first \vb{if} in \refex{breakAndReturn}).

A \it{loop execution} starts from the first time the looping guard is evaluated and ends when the flow of the program continues outside the loop. That is, even if a loop performs no iterations, we consider the single evaluation of the looping guard to false as a loop execution.
The \it{iteration record} of a loop execution is the number of times the looping guard is evaluated to true during the loop execution.

A \it{door-door execution} is a loop execution which ends because the evaluation of the looping guard returns false. The term is coined from the analogy of entering and exiting a room through a door, the conventional entrance into a room. 
On the contrary, a \it{door-window execution} is a loop execution which ends after executing a statement from within the loop body. 
In this case, the termination of the loop can have three causes: the evaluation of a break statement, the evaluation of a return statement, or the raise of an uncaught exception. 
Following the same analogy, the name suggests an unanticipated evacuation from a room throughout a window. 
The \it{exit nature} of a loop refers to the way a loop execution ends: conditional (door-door), break, return or throw exit (door-window).

An \it{infinite execution} happens when a loop never halts; that is, when the looping guard keeps evaluating to true. We refer to \qt{infinite loop} and \qt{infinite execution} interchangeably. 
An \it{idempotent loop} is a loop whose looping guard can be evaluated arbitrarily more times than needed without changing the output of the algorithm. That is, in this type of loops, the correcteness of the loop is defined as a lower bound on the number of iterations.  The loop in \refex{idempotent} illustrates this phenomenon, in a binary search algorithm. If the looping guard is changed --maintaining the loop body intact-- so that the loop performs a linear number of iterations, the output result of the algorithm wouls still be correct.

\rsubsection{Fault class}{theoryBugs}

The fault class we address is ``infinite loop''.
An infinite loop is the infinite repetitive execution of the loop body.
An infinite loop happens when the execution of the loop body does not change anymore the part of the execution state that is observed by the looping condition. 
An infinite loop is critical because:
1) the program is not responsive anymore;
2) the infinite loop consumes 100\% of the CPU on the machine where it happens.

Namely, there are two kinds of infinite loops, related to the two aforementioned roles of a loop condition: \it{wrong precondition} or \it{wrong exit condition}. In the first case, the bug occurs because the program does not skip the loop when it should. In the second case, the bug occurs because the loop does not terminate at the appropriate moment.

To fix a wrong precondition bug, there are two possible repairs.
First, one con wrap the loop  within an \vb{if/then} statement encoding the precondition.
Second, one can modify the loop condition so that the precondition becomes correct while the exit predicate is still valid.

For a wrong exit condition bug, there are three possible repairs:
1) changing the loop condition;
2) adding a window exit such as \qvb{if(X) break} or \qvb{if(X) return};
3) changing the loop body such that the body correctly modifies the execution state that is analyzed in the loop condition.
The automatic repair technique we present in this paper targets a change in the loop condition, which is able to both fix incorrect preconditions and incorrect exit predicates.

\rsection{Contribution}{ourApproach}

In this section we present our approach to fixing infinite loops, called $Infinitel$. 
We focus on  \vb{while} loops where the bug lies in the loop condition. 
According to the analysis presented in \refsection{loopTheory}, our approach repairs wrong exit condition of door-door loop executions. 
Our technique is based on test cases, the infinite door-door executions to be fixed are those manifested while running the test suite. 

In this context, \qt{repairing} the infinite loop means finding a looping guard for the infinite loop such that each test case using that loop both halts and passes all the assertions. We first introduce the overview of our repair approach, and then we proceed by describing each step individually.

\subsection{Overview}

\ba
\Procedure{infiniteLoopRepair}{$src$, $tests$}
\State $src2 \gets \Call{instrumentLoops}{src}$
\State $loop \gets \Call{detectInfiniteLoop}{src2, tests}$
\State $thresholds \gets \Call{findThresholds}{loop, src2, tests}$
\State $patch \gets \Call{findPatch}{loop, src, tests, thresholds}$
\State \Return $patch$
\EndProcedure
\ea{Top level algorithm to repair an infinite loop}{topLevel}

In \refalg{topLevel}, we present the top level algorithm of our repair method. The input for our algorithm is the source code containing an infinite loop (parameter $src$) and the test suite of the source code (parameter $tests$). 
The test suite is composed of passing tests and at least one hanging test, the one that triggers the infinite loop.

The first step is to instrument the source code of the input project $src$. The instrumentation enables us to remotely control loop executions (for instance, to stop tests from hanging). Once the instrumentation is performed, the second step is to detect the presence of an infinite loop during the execution of the test suite. We do this by running the test suite and detecting hanging tests.

In the third step, the goal is to find the number of iterations needed by the detected infinite loop to pass the assertions executed at the end of hanging tests. We call this number a ``threshold'' for that loop. When breaking the infinite loop beyond the threshold, the test case passes.

In the last step, a new looping guard is synthesised, this is the final patch. The detailed explanation of each step is given in the following sections (\refsubsection[Subsections]{repairInstrumentation}, \refsubsection[]{detection}, \refsubsection[]{thresholds} and \refsubsection[]{synthesis}).

\rsubsection{Project Instrumentation}{repairInstrumentation}

\begin{example}
\centering
\begin{minipage}{0.8\linewidth}
\begin{batim}[numbers=left, xleftmargin=8pt]
int method(int a) \{
  int b = a;
\diffpos{LoopMonitor LM\_83 = Global.getMonitor(83);}
\diffpos{int ITERS\_83 = 0;}
\diffneg{while (b > 0) \{}
\diffpos{while (true) \{}
\diffpos{  boolean stay = LM\_83.decide(b > 0, ITERS\_83);}
\diffpos{  LM\_83.collect(stay, b, a, ...);}
\diffpos{  if (stay) \{}
\diffpos{    ITERS\_83 ++;}
      if (b == 18) \{
        return a;
      \}
      if (b == 9) \{
        break;
      \}
      b -=1;
\diffpos{  \} else break;}
  \}
  return b;
\}
\end{batim}
\end{minipage}
\captionexample{Illustration of our loop instrumentation on \refex{breakAndReturn}. The code prefixed by +, in green, is automatically injected with source code transformation.}{approachInstrumentation}
\end{example}

We explain here how to modify the implementation of a loop in order to control its execution. The idea is to \qt{implant} a hook in the loop source code to modify the semantics of the loop at runtime. Specifically, we want to control the looping guard.
The loop instrumentation is shown in \refex{approachInstrumentation}. Firstly, we fetch the loop monitor who will control the loop (line 3). Secondly, a local variable is created to store the iteration record of each loop execution (line 4). Then, we modify the original loop by wrapping the original loop body (lines 11-17) within an if statement (lines 9-18). The original looping guard is deleted (line 5) replaced by a \texttt{while(true)} (line 6). Now, the decision to proceed with a new iteration or break the loop is delegated to the loop monitor. The decision to keep executing the loop or to break is stored in another local variable (line 7). Then, according to this decision, either a new iteration is carried out (\vb{then} branch of the new wrapping \vb{if}) or the loop breaks (\vb{else} branch). In the former case, the local variable is incremented (line 10). Finally, we add one more statement on this instrumentation to collect the execution information of each iteration (line 8).

\rsubsection{Infinite Loop Detection}{detection}

\ba
\Procedure{detectInfiniteLoops}{$src$, $tests$}
\State $hangingTests \gets \{\}$
\State $monitors \gets \Call{implantedMonitors}{\null}$
\State \Call{setLimitInAll}{$monitors$, $\nat{1000000}$}
\For{$test \in tests$}
	\State $\Call{run}{src, test}$
	\For{$monitor \in monitors$}
		\If{$monitor.\Call{hasExceedingExecution}{\null}$}
		    \State $loop \gets monitor.\Call{getLoop}{\null}$
      		\State $invocation \gets monitor.\Call{getExceedingExecution}{\null}$
			\State $hangingTests.\Call{put}{test, loop, invocation}$
		\EndIf
	\EndFor
\EndFor\\
\Return $hangingTests$
\EndProcedure
\ea{Detecting infinite loops with instrumentation.}{detection}

Our method to detect infinite loops is straightforward. We keep track of the number of iterations throughout a loop execution and, if a maximum number of iterations is exceeded, we assume it is an infinite loop. 
We implement this detection strategy with the non-trivial instrumentation explained in \refsubsection{repairInstrumentation}.

During the loop execution, the loop monitor is responsible for deciding whether to iterate or break before starting a new iteration. To do this, it receives the evaluation of the original looping guard and the number of already completed iterations. If this number exceeds a maximum number of iterations, the loop monitor labels the loop as ``infinite'' and breaks it. 
The threshold is fully parameterizable, we use a reasonable value of 1 million. Across all our experiments, this has only yielded one false positive (wrongly detected as infinite loop). 

The infinite loop detection is detailed in \refalg{detection}. At this stage, the parameter $src$ is the instrumented source code and the parameter $tests$ is the test suite. We simply run the whole test suite on $src$. Every loop execution is monitored by a loop monitor. In the event of an infinite execution of a hanging test, the loop monitor will detect the infinite execution and it will break the loop  when the threshold is exceeded. Also, because this infinite loop is detected during the infinite execution, the loop monitor stores the invocation rank of the infinite execution (for instance, \qt{the fourth loop execution of hanging test \vb{testABC}}).
The output of this algorithm is a specific data structure that contains  the list of hanging tests, the infinite loop where each one hangs, and the invocation rank of the infinite execution in each case.

\rsubsection{Finding Thresholds in Hanging Tests}{thresholds}

\ba
\Procedure{findThresholds}{$loop$, $src$, $tests$}
\State $thresholds \gets Dictionary.\Call{new}{\null}$
\State $monitor \gets loop.\Call{getMonitor}{\null}$
\State $hangingTests \gets \Call{hangingTestsOf}{loop}$
\For{$test \in hangingTests$}
    \State $number \gets \Call{getExceedingExecution}{test, loop}$
    \For{($i = 0$; $i \leq \nat{1000000}$; $i++$)}
        \State $monitor.\Call{setLimitIn}{number, i}$
        \State $result \gets \Call{run}{src, test}$
        \If{$result.\Call{isSuccesful}{\null}$}
        \State $thresholds.\Call{put}{test,i}$
        \State \b{break}
        \EndIf
    \EndFor
\EndFor
\EndProcedure
\ea{Identifyig angelic record	s}{findThresholds}

A hanging test, when executed, gets trapped in an infinite loop execution because the looping guard never evaluates to false. That is, the looping guard does not break the loop when it should. To rectify this, we have to amend the looping guard so that it breaks the loop during the infinite execution at the \it{appropriate} moment. 
Hence, we first have to determine the appropriate moment to break the loop in each infinite execution.

We estimate the appropriate moment to break the loop in an infinite execution by controlling the iteration record of the infinite loop in that execution. As seen in \refsubsection{detection}, we can use the loop monitor to break any loop by simply setting a maximum value of permitted iterations. If we set the maximum value equal to $\chi$ during the infinite execution of an infinite loop, and we observe that the hanging test both halts and passes, then we have found this appropriate moment, it's when $\chi$ iterations have been executed.
We refer to the target $\chi$ value as an \qt{angelic record}.
We use this terminology based on the literature terminology (\cite{angelic}, \cite{nopol}). 

Our method to find the angelic record of a hanging test is simple: we explore values from 0 to the predefined threshold in order, run the hanging test each time and assess whether it passes. If it does, the probed value is the angelic record $\chi$. The rationale of this simple strategy is that we have observed that in real test suites, the number of loop iterations is likely a low value (less than 20). Therefore, not only we expect the angelic record to be within that range (0 and 1 million), but, also, we know that probing in ascending order is the fastest way to find the angelic record. 

The angelic record search is detailed in \refalg{findThresholds}. We receive a detected infinite loop (parameter $loop$), the instrumented source code (parameter $src$) and the test suite of the project (parameter $test$). From the previous step (\refsubsection{detection}), we already know the hanging tests of an infinite loop.
 Then, for each hanging test, we probe different values until we find the angelic record. We do this for all hanging tests. We store this information in an associative array where the key is an infinite loop under repair.

\rsubsection{Patch Synthesis}{synthesis}

To synthesize a new looping guard, we use a program synthesis technique. The idea is to synthesize a new looping guard that would make all test passing.

\rsubsubsection{Synthesis as an SMT Problem}{smtIntro}

In this section we briefly introduce the code synthesis method to be used in this paper. The goal of code synthesis is to synthesize a program $\Psi$ which complies with a specification.
Input-output synthesis is one kind of synthesis: for any specified input $\tilde{I}$, the synthesized program $\Psi$ should output an acceptable output $\tilde{O}$. To synthesize $\Psi$, the code synthesis algorithm receives an \qt{input-output} pair set $\mathcal{V}$. For any given pair $(I,O) \in \mathcal{V}$, whenever $I$ is the argument of $\Psi$, then the program has to return the value $O$.

We use \it{component-based synthesis} \cite{oracleGuided}. In addition to input-output pairs, this synthesis algorithm takes a set of \qt{base components} $\mathcal{C}$ (operators such as $>$). For any given pair $(I, O) \in \mathcal{V}$, whenever $I$ is the argument of $\Psi$, then the program has to return the value $O$; and, in order to compute this value, it must only use operators from the $\mathcal{C}$ set.

Component-based synthesis encodes $\mathcal{C}$ and  $\mathcal{V}$  as first-order logic constraints. The constraints both describe the syntax of the algorithm (such as number of lines, declaration of local variables, etc) and the semantics (to make the program compliant with the specification). Then, an SMT solver is used to decide whether there exist a solution satisfying all constraints. If there is, the solution is decoded back and translated into an algorithm.

\rsubsubsection{Synthesis of a New Looping Guard}{ourSynthesis}

We now use the synthesis method described in \refsubsubsection{smtIntro} to generate a new looping guard for the infinite loop. The specification of the new looping guard can be informally expressed as follows: the looping guard predicate should allow every test executing the infinite loop to both halt and pass.

To synthesize the new looping guard, we need two arguments: the input-output pair set and the component set. We first explain how to yield the latter, and then the former.

\textbf{Component Set:}
The component set used for synthesis contains different kinds of operators.
In $Infinitel$,  we use comparison operators ($C_{>}$, $C_{\geq}$, $C_{=}$, $C_{\neq}$), then logic operators ($C_{not}$, $C_{or}$, $C_{and}$), then linear arithmetic operators ($C_{+}$, $C_{-}$), then \vb{if-then-else} ($C_{ite}$), and, finally, multiplication ($C_{\times}$).

The selection of components is done increasingly. We start off with an empty set of components. In this case, there is only one possibility to find a new looping guard: the patch uses a boolean input variable. That is, an input $i \in I$ of the specification is equal to the corresponding output for every $(I,O)$ pair of $\mathcal{V}$. If this is the case, then the new looping guard is simply \qvb{while (i)}. 

If not, we formulate a new SMT problem with the same specification $\mathcal{V}$ and a new non-empty component set. If we succeed, the synthesis phase is finished. If not, we keep adding components until either a new looping guard is found, or until we exhaust all of the available components and finish the synthesis phase unsuccessfully.

\textbf{Input-Output Pair Set:} The input-output specification $\mathcal{V}$ is assembled as follows. As explained in the designed project instrumentation (\refex{approachInstrumentation}), the loop monitor is a reification of the looping guard: it decides to iterate or break the loop in each iteration. 
The collection of input-output pairs is done by the loop monitor.
At each iteration it creates an $(I,O)$ pair associating the decision of the loop monitor to $O$ and the context information to $I$, whose collection is described in \refsubsubsection{runtimeCollection}.

\ba
\Procedure{specification}{$loop, thresholds, src$}
\State $V \gets Set.\Call{new}{\null}$
\State $monitor \gets loop.\Call{getMonitor}{\null}$
\State $tests \gets \Call{testsOf}{loop}$
\For{$test \in tests$}
	\If{$\Call{isHangingTestOf}{loop, test}$}
		\State $number \gets \Call{getExceedingExecution}{test, loop}$
		\State $threshold \gets thresholds.\Call{keyFor}{test}$
		\State $monitor.\Call{setLimitIn}{number, threshold}$
	\EndIf
	\State $\Call{run}{src, test}$
	\State $pairs \gets monitor.\Call{getPairs}{\null}$
	\State $V.\Call{addAll}{pairs}$ 
\EndFor
\EndProcedure
\ea{Obtaining the input-output pair set}{specification}

\rsubsubsection{Runtime Value Collection}{runtimeCollection}

The context information is collected by the loop monitor within the  callback in line 8 of \refex{approachInstrumentation}. 
The context information reflects the local state of the program at each iteration. It is composed of variables collected in 6 different ways: 

\disableindent

\it{Reachable variables hunt:} we scan the scope of the loop to gather every \it{reachable variable}. A reachable variable is a variable with two qualities: it is accessible from the loop scope (it is declared within the lexical scope of the loop) and it is initialized. It could either be a local variable, a method parameter or an instance field.

\it{Visible field access:} for each reachable variable of a user-defined type, we also gather its visible fields. 

\it{Getters:} for a reachable variable of a user-defined type, we also include, when possible, non-visible fields. To do this, we review the source code declaration of the variable class in search of \qt{getter methods}. A getter method is a method with the following characteristics: it has no parameters, it is implemented in one line, the line is a \vb{return} statement, and the returned element is an instance field. 

\it{Recycling of the original looping guard:} although the original looping guard is not used as the real looping guard after the loop instrumentation, it is highly likely that it still provides precise information about the iteration context. For this reason, we also include the value of the evaluation of the original looping guard.

\it{Subvalues of the original looping guard:} whenever possible, we also inspect the values of subcomponents of the original looping guard. For instance, if the original looping guard is a conjunction, we also include the evaluation of each subpredicate of the conjunction.

\enableindent

In essence, we have many options to gather context information within the lexical scope of the loop. However, only boolean or numeric values are supported by SMT solvers. For this reason, we need to further refine the amassed variables:\\

\disableindent

\it{Extraction by value:} for a variable of primitive type (\vb{boolean}, \vb{char}, \vb{int}, \vb{double}, etc.) we take its value.\\

\it{Extraction by queries:} for each gathered variable of a non-primitive type, we perform different queries. First, we check nullness; and, if the variable is not \vb{null}, we also extract information by using a hardcoded list of typical queries (such as the length of a \vb{String}, or the size of a \vb{List}). We can see the hardcoded queries in \reftable{typeQueries}. The way to interpret the table is: if the variable's class subclasses a given superclass, then we perform the corresponding queries on the variable.

\enableindent

\bt{c|v}
\linerowlinespace{\b{Superclass} & \multicolumn{1}{c}{\b{Queries}}}
\linerowlinespace{Object & variable != null}
\linerowlinespace{Array & variable.length}
\linerowlinespace{Iterator & variable.hasNext()}
\linerowlinespace{Enumeration & variable.hasMoreElements()}
\linerow{\multirow{2}{*}{Collection (incl. Lists and Maps)} & variable.size()} \cline{2-2}
\rowlinespace{& variable.isEmpty()}
\et{Queries for each type}{typeQueries}

The last step consists of a refinement of the input-output pair set, in order to improve the SMT synthesis. Firstly, we enrich each set $I$ of every $(I,O)$ pair with the constant values of $-1$, $0$ and $1$. These are values commonly used in predicates, so we make sure they are available for code synthesis. Then, we remove input elements which have the same value in every $(I,O)$ pair.

\subsubsection{Synthesis Algorithm}

\ba
\Procedure{findPatch}{$loop, thresholds, src, tests$}
\State $spec \gets \Call{specification}{src, loop, thresholds}$
\State $components \gets List.\Call{new}{\null}$
\While{\b{not} $\Call{exhaustedAll}{components}$}
    \State $smtProblem \gets \Call{encodeToSMT}{spec, components}$
    \If{$smtProblem.$\Call{isFeasible}{\null}}
        \State $solution \gets smtProblem.$\Call{solution}{\null}
        \State $patch \gets \Call{decodeToPatch}{solution}$
        \State \b{break}
    \EndIf
	\State $bundle \gets \Call{nextComponentBundle}{\null}$
	\State $components.\Call{addAll}{bundle}$
\EndWhile
\EndProcedure
\ea{Patch synthesis}{synthesis}

The algorithm of the new looping guard synthesis can be seen in \refalg{synthesis}. The first step is to collect the input-output pair set (detailed in \refalg{specification}). To do this we simply execute each test using the loop and fetch the collected input-output pairs after each run. 
For hanging tests, the expected output is based on the found infinite execution thresholds. Once we obtain this specification, we start the search of a new looping guard. We begin with an empty component set, we formulate an SMT problem and use a solver to find a solution. If we succeed, we transform back the SMT solution into a boolean code expression, the patch. If not, we add components to the component set and formulate a new SMT problem. We do this until we exhaust all of the components or a correct looping guard has been synthesized.  

To explain the details of this synthesis technique, it requires a large amount of space. Since it is not the contribution of this paper, we refer the reader to the original paper \cite{oracleGuided} and our paper presenting our adaptation of the technique to handle object-oriented code \cite{nopol}.

\rsection{Evaluation}{evaluation}

In this section we present the evaluation of $Infinitel$, our system for automatically repairing infinite loops. Our evaluation is based on the repair of 7 seeded bugs and 7 real bugs. We aim to answer the following research questions:

\be[label=\it{RQ\arabic*}]

\labelleditem{performance} \it{Performance:} does $Infinitel$ solve the bugs in a reasonable amount of time? What is the bottleneck of the repair method? 

\labelleditem{correctness} \it{Appropriateness:} are the patches synthesized by $Infinitel$ appropriate?  How do they compare with the human-written ones?

\labelleditem{adequacy} \it{Synthesis:} How hard is the code synthesis for each bug? does the synthesis based on SMT satisfiability scale?

\labelleditem{limit} \it{Angelic Record (\refsubsection{detection}):} does the 1 million iteration limit used for finding the angelic record have type I errors (false positives)? Does it affect the performance of $Infinitel$?

\labelleditem{technique} \it{State Observation (\refsubsubsection{runtimeCollection}):} how does our value collection technique qualify the loop state in each bug?

\labelleditem{idempotence} \it{Idempotence:} is there any idempotent loop among the real bugs? 

\ee

\subsection{Evaluation Setup}

$Infinitel$ is implemented in $Java$, running on an Oracle JRE version 7, with a maximum heap of 2 GB. The SMT solver used is $Z3$\footnote{\url{http://z3.codeplex.com/}} version 4.3.2. The operating system where the evaluation is performed is \vb{OS X Mavericks}.  

\subsubsection{Methodology}

Our evaluation is based on the repair of 7 seeded bugs and 7 real bugs. In both cases, each bug consists of one infinite loop with at least one hanging test. 
The magic number 7 comes from the fact that we were able to reproduce 7 real bugs within the 6 weeks we allocated for bug reproduction, and we wanted the same number for symmetry in the result tables.

\subsubsection*{Seeded Bugs}

A seeded bug is a project deliberately \qt{infected} with a manually created infinite loop. We first create four toy projects  which only have one class, one test class and one infinite loop. They are called  Ex \{1\ldots4\}.
 
In addition, we seed another 3 infinite loops in large-scale open-source projects. 
Those seeded bugs are more representative than the toy projects because they are at real scale. While the bugs are artificial, the meaning of performance metrics are meaningful for those cases.
The process is as follows, we select a loop in the project and we perform two small transformations on it. Firstly, we substitute the looping guard with \qvb{while (true)}\footnote{Actually, changing to \qvb{while (true)} would raise compilation errors because of the presence of unreachable code after the loop. Hence, we use an equivalent form: \qvb{while ("".isEmpty())}.}. Secondly, the loop body is wrapped with a \vb{try/catch} with an empty \vb{catch} block. This is done to prevent an exception from breaking the infinite loop.

For the large scale evaluation seeded bugs we use Apache\rq{}s Commons $Collections$ and $Math$ projects (commits \vb{b5ffdaf} and \vb{32ef444}). The first two seeded bugs are in two different loops on $Collections$ (\vb{AbstractMapBag.java} on line \nat{590} and \vb{AbstractDualBidiMap.java} on line 352), whereas the third one comes from an infected loop on $Math$ (\vb{FastMath.java} line \nat{3120}).

\subsubsection*{Real Bugs}

\bt{ *{5}{c} }
\hline
\rowline{Name & Repository & Commit & Subproject & Test}
\row{csv & \url{git://git.apache.org/commons-csv.git} & 4dfc8ed & -- & \vb{Y}}
\row{fop & \url{git://git.apache.org/fop.git} & 13984cc & -- & \vb{N}}
\row{pdfbox A & \url{git://git.apache.org/pdfbox.git} & b10cf48 & -- & \vb{N}}
\row{pdfbox B & \url{git://git.apache.org/pdfbox.git} & a2ab77f & fontbox & \vb{N}}
\row{pig & \url{git://git.apache.org/pig.git} & 5abfbd0 & piggybank & \vb{Y}}
\row{tika & \url{git://git.apache.org/tika.git} & 1b694e7 & tika-parser & \vb{N}}
\rowline{uima & \url{git://git.apache.org/uima-uimaj.git} & 155596a & jVinci & \vb{N}}
\et{Our dataset of 7 real bugs, up to the commit ID (the first 7 digits of the commit checksum)}{projectList}

The seven real bugs come from existing projects of the Apache Git repositories\footnote{\url{http://git.apache.org/}}. To find real bugs, we have individually analyzed the projects looking for commits reporting and fixing an infinite loop bug. Specifically, we perform a keyword-based search on the Git\footnote{\url{http://git-scm.com/}} log of each project repository (keywords: infinite, loop, iteration, hang, endless, ending, terminating).

We describe each real bug in \reftable{projectList}.  In the case of \vb{csv} and \vb{pig} the commit includes code changes to fix the infinite loop and a test case validating those changes (i.e. triggering the infinite loop). For the rest of the commits, the test cases triggering the infinite loop are missing. Consequently, we manually created tests for the 5 remaining commits. The policy followed to manually create tests is the following: \it{a}) at least one of these tests has an infinite execution of the loop attempted to be fixed by the commit changes; \it{b}) the added hanging tests halt and pass with the changes introduced in the commit; and, \it{c}) the added and not hanging tests, if any, pass.
(We should mention that the \vb{pdfbox B} bug is detected and incorrectly reported as fixed in commit \vb{e41cbd1}, but it is actually  fixed in later commit \vb{a2ab77f}. We use the buggy loop in the first commit and use the second commit to compare the synthesized looping guard with the manually written fix.)

Note that it takes a lot of time to collect and reproduce real bugs of a given defect class. For those 7 bugs, it took us more than 6 weeks.
For sake of comparison, the close related work on infinite loops use respectively  eight bugs \cite{jolt} and one single bug \cite{Burnim2009,ibing2015fixed} in their evaluation.

\subsubsection{Metrics}

We now present the different evaluation metrics about the automatic repair of each bug. We group them in two categories: \qt{basic metrics} and \qt{time metrics}. 
Unless indicated otherwise, each time metric is rounded to seconds. Time metrics are used to answer \refitem{performance}.

\subsubsection*{Basic Metrics}

\disableindent

\it{Tests}: the total number of tests in the test suite of the project.

\it{Application Classes:} the total number of declared classes in the project source code, excluding test classes. This metric is also equal to the total number of classes which are instrumented and recompiled during the project instrumentation (\refsubsection{repairInstrumentation}).

\it{Added Tests:} the number of tests added to reproduce the infinite loop bug (only in real bugs).

\it{Added LOC:} the total number of lines of the added tests (only in real bugs).

\it{Hanging Tests:} the number of invoking tests which do not halt due to the infinite loop. 

\it{Idempotence:} whether the hanging tests pass or not when they are run with an arbitrary large number of executions. If they do, we suspect that the infinite loop behaves like an idempotent loop (see \refsection{loopTheory}).
 
\it{Angelic Record:} the value of the highest angelic record for the infinite loop under repair.

\it{Context Items:} the size of the input-output pair set described in \refsubsubsection{ourSynthesis}. This number impacts the number of constraints in the SMT problems created during code synthesis.

\it{Context Size:} the number of inputs inside each input-output pair, plus 1 (for the output value). It represents the number of extracted values being used to describe the state of each loop iteration (\refsubsubsection{runtimeCollection}).

\it{SMT Formulations:} the number of total SMT problems needed to synthesize a patch. As indicated in \refalg{synthesis}, we successively create SMT problems by adding new components until the synthesis succeeds. The number of SMT problems needed to find a solution is a proxy to the number and complexity of the components used for synthesis, it enables us to qualify the difficulty of the found patch.

\it{SMT Components:} the number of total components used in the synthesized patch (\refsubsubsection{smtIntro}).

\it{SMT Component Types:} the number of different component types used in the synthesized patch (there are 5 different types: comparison, logic, linear arithmetic, multiplication and \vb{if-then-else}).

\it{LOC:} lines of code in the project source code, excluding test code. Figures are obtained with $CLOC$\footnote{\url{http://cloc.sourceforge.net/}}.

\enableindent

\subsubsection*{Time Metrics}

\disableindent

\it{Instrumentation:} time to implant the loop monitors in every \vb{while} of the project source code (\refsubsection{repairInstrumentation}).

\it{Compilation:} time to compile the instrumented source code.

\it{Test Suite:} time to run the test suite of the project. This metric includes the time of running --and inducing loop termination of-- hanging tests.

\it{Hanging Tests:} time to run hanging tests of the infinite loop by breaking after the maximum number of iterations. Every infinite execution is interrupted after a maximum iteration number is reached (\refsubsection{detection}).

\it{Angelic Value Mining:} time to find the angelic records of each hanging test (\refsubsection{thresholds}).

\it{Value Collection:} time to collect contexts for tests invoking the infinite loop (\refsubsubsection{runtimeCollection}).

\it{SMT Solving:} overall time solving all SMT problems until a solution is found.

\it{Total Time:} the sum of the previous 7 metrics, it is the total execution time to automatically fix the bug.

\enableindent

\newcolumntype{P}{>{\raggedleft\arraybackslash}p{1.2cm}} 

\bt{ l | *{6}{P|} r} 
\fullheader{8}{Basic Metrics}
\tablespace

\linerow{ & \multicolumn{2}{c|}{\bf{Collections}} & & & & &} \cline{2-3}
\rowline{\tabhead{c|}{Seeded} & \tabhead{c|}{A} & \tabhead{c|}{B} & \tabhead{c|}{Math} & \tabhead{c|}{Ex.1} & \tabhead{c|}{Ex.2} & \tabhead{c|}{Ex.3} & \tabhead{c}{Ex.4}}
\row{LOC & \nat{25338} & \nat{25338} & \nat{91878} & \nat{11} & \nat{10} & \nat{15} & \nat{43}}
\row{Application Classes & \nat{463} & \nat{463} & \nat{1188} & \nat{1} & \nat{1} & \nat{1} & \nat{1}}
\row{Tests & \nat{14792} & \nat{14792} & \nat{6077} & \nat{5} & \nat{2} & \nat{3} & \nat{3}}
\row{Hanging Tests & \nat{11} & \nat{57} & \nat{1} & \nat{1} & \nat{1} & \nat{1} & \nat{1}}
\row{Context Items & \nat{3} & \nat{3} & \nat{53} & \nat{15} & \nat{4} & \nat{3} & \nat{61}}
\row{Context Size & \nat{15} & \nat{9} & \nat{67} & \nat{4} & \nat{8} & \nat{4} & \nat{9}}
\row{SMT Formulations & \nat{1} & \nat{1} & \nat{2} & \nat{4} & \nat{1} & \nat{2} & \nat{2}}
\row{SMT Components & \nat{0} & \nat{0} & \nat{1} & \nat{9} & \nat{0} & \nat{1} & \nat{1}}
\row{SMT Comp. Types & \nat{0} & \nat{0} & \nat{1} & \nat{3} &  \nat{0} & \nat{1} & \nat{1}}
\rowlinespace{Angelic Record & \nat{0} & \nat{18} & \nat{52} & \nat{4} & \nat{1} & \nat{0} & \nat{37}}

\linerow{ & & & \multicolumn{2}{c|}{\bf{pdfbox}} & & & } \cline{4-5}
\rowline{\tabhead{c|}{Real} & \tabhead{c|}{csv} & \tabhead{c|}{fop} & \tabhead{c|}{A} &\tabhead{c|}{B} & \tabhead{c|}{pig} & \tabhead{c|}{tika} & \tabhead{c}{uima}}
\row{LOC & \nat{1218} & \nat{157445} & \nat{39551} & \nat{9339} & \nat{11380} & \nat{19767} & \nat{7135}}
\row{Application Classes & \nat{11} & \nat{2340} & \nat{429} & \nat{93} & \nat{211} & \nat{263} & \nat{83}}
\row{Tests & \nat{83} & \nat{2693} & \nat{24} & \nat{11} & \nat{243} & \nat{476} & \nat{2}}
\row{Hanging Tests & \nat{1} & \nat{3} & \nat{2} & \nat{1} & \nat{1} & \nat{1} & \nat{1}}
\row{Context Items & \nat{6631} & \nat{209} & \nat{1474} & \nat{6} & \nat{23} & \nat{703} & \nat{12}}
\row{Context Size & \nat{41} & \nat{15} & \nat{15} & \nat{11} & \nat{14} & \nat{37} & \nat{13}}
\row{SMT Formulations & \nat{2} & \nat{4} & \nat{2} & \nat{3} & \nat{3} & \nat{3} & \nat{2}}
\row{SMT Components & \nat{1} & \nat{4} & \nat{1} & \nat{2} & \nat{2} & \nat{4} & \nat{1}}
\row{SMT Comp. Types & \nat{1} & \nat{3} & \nat{1} & \nat{2} & \nat{2} & \nat{2} & \nat{1}}
\row{Angelic Record & \nat{1} & \nat{45} & \nat{0} & \nat{0} & \nat{3} & \nat{0} & \nat{10}}

\hline
\row{Added Tests & \nat{0} & \nat{3} & \nat{1} & \nat{4} & \nat{0} & \nat{1} & \nat{2}}
\row{Added LOC & \nat{0} & \nat{36} & \nat{25} & \nat{32} & \nat{0} & \nat{22} & \nat{11}}
\rowline{Idempotence & \vb{Y} & \vb{Y} & \vb{Y} & \vb{Y} & \vb{Y} & \vb{N} & \vb{N}}

\et{Evaluation of \it{Infinitel} on our Dataset}{results}

\subsection{Empirical Results}

We answer the research questions presented in \refsection{evaluation}.

\subsubsection{Descriptive Statistics} 
\reftable{results} summarizes all the evaluation metrics.
The table is composed of two parts.
The upper part is about the 7 seeded bugs, the lower part is about the 7 real infinite loops.

For each of them, the first lines give descriptive statistics, 
the number of lines of code,
the number of application classes (excluding test classes),
the number of test methods (the basic test unit in the testing framework JUnit),
and the number of test methods that hang because of the infinite loop.
For the seeded bugs, one can see that the first three infinite loops have been seeded  in large projects: the version of Apache Commons Collections under repair, called ``Collections A'', has \nat{25 338} lines of code, over 463 classes.
The test suite specifies \nat{14 792} test case.
The seeded infinite loop breaks between 1 (for Math, Ex \{1\ldots4\}) and 57 test cases (for Collections B).

For the real bugs, the projects range between \nat{1218} LOC (for csv) and \nat{157 445} LOC for fop.
The number of applications classes and test cases follow the same trend.
For 5/7 real bugs, there is one single test case that triggers the infinite loop, however, there are 3 (resp. 2) hanging tests for fop (resp. pdfbox A).

\subsubsection{\refitem{performance} Performance} 
Our automatic repair algorithm automatically fixes all seeded and real bugs.
Let us now discuss the performance of the technique.
Does $Infinitel$ solve the bugs in a reasonable amount of time? What is the bottleneck of the repair method? 
For the seeded bugs, the interesting cases are the bugs put in large-scale projects.
Looking at the time metrics in \reftable{performance}, we notice that the total execution time for the Apache projects is around 10 minutes.
In all three cases, the bottleneck is the running time of the test suite.
In the case of \vb{Collections B}, the hanging tests account for approximately \percent{50} of the time of running the test suite (before breaking after 1 million iterations) the reason is that the seeded bug in \vb{Collections B} creates 57 hanging tests.

We now look at the performance metrics for the real bugs in the bottom part of \reftable{performance}.
We can see that \vb{fop} has the longest repair time with approximately 1 hour. Then comes \vb{uima} with almost 49 minutes and \vb{csv} with roughly 30 minutes. For the other 4 real bugs, 10 minutes is enough to repair them. We now analyze the bottleneck of each bug individually.

In \vb{csv} the clear bottleneck is the time of SMT Solving: \percent{99} of the total repair time is spent in that task. Something similar happens in \vb{fop}, with \percent{88}. This is due to the size of the SMT problems in \vb{csv} (where each of the \nat{6631} context items accounts for a constraint in the SMT problem) and to the complexity of the found patch in \vb{fop} (it requires many components as witnessed by the highest number of SMT problems: 4 -- which means that new component types were added four times in a row).

In \vb{uima}, running the hanging tests is the evident bottleneck. This is due to a performance overhead caused by the string concatenation operation. The infinite loop in \vb{uima} only has one statement which is a concatenation of strings with the sum operator; consequently, for each of the  one million iterations before the forced break, two strings are created and copied.

In \vb{pdfbox A} and \vb{pig}, the bottleneck seems again to be related to the test cases. In \vb{pdfbox A} almost \percent{70} of the time to repair the infinite loop (84 seconds) is spent for executing the test suite (57 seconds) or collecting the values at runtime (27 seconds). This figure increases to \percent{95} of the time (503 seconds) in \vb{pig} (442 seconds running the test Suite and 61 seconds for Value Collection).

In \vb{tika} the combined time for running the test suite and the SMT solving amounts for more than \percent{90} of the total repair time. Finally, in \vb{pdfbox B} the SMT solving is negligible and in this case, the repair time is dominated by the project instrumentation. 

To sum up, according to our dataset, $Infinitel$ is able to fix infinite loops on a standard laptop computer. 

\bt{ l| *{6}{P|} r} 
\fullheader{8}{Time Metrics (in seconds)}
\tablespace

\linerow{ & \multicolumn{2}{c|}{\bf{Collections}} & & & & &} \cline{2-3}
\rowline{\tabhead{c|}{Seeded} & \tabhead{c|}{A} & \tabhead{c|}{B} & \tabhead{c|}{Math} & \tabhead{c|}{Ex.1} & \tabhead{c|}{Ex.2} & \tabhead{c|}{Ex.3} & \tabhead{c}{Ex.4}}
\row{Instrumentation & \timesec{17.588} & \timesec{17.273} & \timesec{34.112} & \timesec{2.230} & \timesec{1.067} & \timesec{1.024} & \timesec{1.081}}
\row{Compilation & \timesec{13.077} & \timesec{12.258} & \timesec{16.646} & \timesec{1.304} & \timesec{0.634} & \timesec{0.622} & \timesec{0.776}}
\row{Test Suite & \timesec{245.300} & \timesec{794.934} & \timesec{589.395} & \timesec{0.149} & \timesec{0.125} & \timesec{0.251} & \timesec{0.135}}
\row{Hanging Tests & \timesec{47.148} & \timesec{406.958} & \timesec{0.467} & \timesec{0.137} & \timesec{0.113} & \timesec{0.243} & \timesec{0.121}}
\row{Angelic Record Mining & \timesec{13.911} & \timesec{2.117} & \timesec{0.389} & \timesec{0.067} & \timesec{0.007} & \timesec{0.005} & \timesec{0.114}}
\row{Value Collection & \timesec{0.598} & \timesec{1.398} & \timesec{0.032} & \timesec{0.044} & \timesec{0.015} & \timesec{0.014} & \timesec{0.024}}
\rowline{SMT Solving & \timesec{11.422} & \timesec{6.127} & \timesec{6.069} & \timesec{6.204} & \timesec{0.007} & \timesec{0.256} & \timesec{0.638}}
\row{Total Time& \timesec{349} & \timesec{1240} & \timesec{646} & \timesec{9} & \timesec{2} & \timesec{2} & \timesec{3}}
\rowlinespace{Total Time (readable) & \vb{0:05:49} & \vb{0:20:40} & \vb{0:10:46} & \vb{0:00:09} & \vb{0:00:02} & \vb{0:00:02} & \vb{0:00:03}}

\linerow{ & & & \multicolumn{2}{c|}{\bf{pdfbox}} & & & } \cline{4-5}
\rowline{\tabhead{c|}{Real} & \tabhead{c|}{csv} & \tabhead{c|}{fop} & \tabhead{c|}{A} &\tabhead{c|}{B} & \tabhead{c|}{pig} & \tabhead{c|}{tika} & \tabhead{c}{uima}}
\row{Compilation & \timesec{1.452} & \timesec{32.722} & \timesec{6.993} & \timesec{3.256} & \timesec{7.425} & \timesec{13.242} & \timesec{4.014}}
\row{Test Suite & \timesec{1.814} & \timesec{330.068} & \timesec{56.863} & \timesec{0.304} & \timesec{441.742} & \timesec{102.293} & \timesec{2897.701}}
\row{Hanging Tests & \timesec{0.684} & \timesec{0.246} & \timesec{0.780} & \timesec{0.251} & \timesec{69.251} & \timesec{0.522} & \timesec{2897.194}}
\row{Angelic Record Mining & \timesec{0.002} & \timesec{0.989} & \timesec{0.009} & \timesec{0.007} & \timesec{0.057} & \timesec{0.006} & \timesec{0.276}}
\row{Value Collection & \timesec{10.944} & \timesec{4.988} & \timesec{26.527} & \timesec{0.016} & \timesec{60.627} & \timesec{10.837} & \timesec{0.115}}
\rowline{SMT Solving & \timesec{1780.634} & \timesec{3204.781} & \timesec{21.138} & \timesec{0.775} & \timesec{7.164} & \timesec{368.976} & \timesec{3.176}}
\row{Total Time & \timesec{1797} & \timesec{3619} & \timesec{123} & \timesec{7} & \timesec{525} & \timesec{509} & \timesec{2911}}
\rowline{Total Time (readable) & \vb{0:29:57} & \vb{1:00:19} & \vb{0:02:03} & \vb{0:00:07} & \vb{0:08:45} & \vb{0:08:29} & \vb{0:48:31}}
\et{Performance of \it{Infinitel} on our dataset}{performance}

\subsubsection{\refitem{correctness} Appropriateness} 

Are the patches synthesized by $Infinitel$ appropriate?  How do they compare with the human-written ones?
Now we assess the appropriateness of the synthesized looping guards.
For the seeded bugs, for Ex \{1\ldots4\} the patch is obviously the expected one since we craft those examples manually. 
For \vb{Collections B} the original looping guard (before seeding the bug) was restored. 
For \vb{Collections B} and \vb{Math}, the original looping guard is restored semantically, but not syntactically.
In the case of  \vb{Math}, whereas the original looping guard checks that the variable \vb{mantissa} is lower than $2^{52}$ using bitwise right shift operator, the found looping guard does so by comparing with the value of constant \vb{TWO\_POWER\_52}, which is equivalent. 

Let us now concentrate on the real infinite loops.
For each subject, \reftable{realFixes} gives both the synthesized patch and the human fix.
We consider \vb{csv} and \vb{pig} bugs. 
In the case of \vb{pig}, the human fix and the found patch are equivalent (\vb{fileStatusArr} is an array, so the length is either 0 or positive).
For \vb{csv}, despite that the fixes are different, both the human fix and the found patch base the looping guard on the value of \vb{tkn.type}.  According to our understanding of the program, they are equivalent.

We now consider \vb{tika} bug. Our found patch simply restricts the original looping guard with an additional on the block length (first operand of the patch).
On the contrary, the human fix is more complex, because it uses an additional boolean variable \vb{continueLoop}. This new variable is updated at the end of every iteration, and the value of this variable is used in the new looping guard.
This manual code is likely more readable than the synthesized one.

If we analyze \vb{fop} bug, we will again find different repair strategies. This time, the human fix adds a break statement at the end of the loop body. As we have already discussed in \refsubsection{theoryBugs}, a behavior equivalent to adding a break can often be obtained by modifying the looping guard itself, this is what $Infinitel$ does.

The same happens when we compare the found patch and the human fix in bug \vb{pdfbox A} . The human wraps the while with an if statement that acts as a precondition. 
Logically, $Infinitel$ finds that the angelic record for the hanging tests in \vb{pdfbox A} (\reftable{results}) is 0, indicating that the loop body should not be executed at all. Later on in the repair process, $Infinitel$ synthesizes an expression that indeed acts both as a correct precondition and a correct exit condition. 

For the two remaining ones \vb{pdfbox B} and \vb{uima}, alternative strategies are used in the human fix. 
The human fix for \vb{pdfbox B} modifies a method invoked in the looping guard.
The developer patch of \vb{uima} adds a statement to the loop body.

To sum up, there are many alternative strategies to repair an infinite loop.
Listings \refex{csvLoop}, \refex{tikaLoop}, \refex{fopLoop} and \refex{pdfboxLoop} summarizes them.  
This evaluation shows that our repair strategy (modifying the looping guard) is as powerful as the others. Having a single repair strategy enables us to greatly reduce the search space of the patch.

\bt{|c|c|}

\fullheader{2}{csv}
\linerow{faulty & \vb{!tkn.isReady}}
\linerow{manual & \vb{!tkn.isReady \&\& tkn.type != TT\_EOF}}
\linerowlinespace{infinitel & \vb{(tkn.type)<(0)}}

\fullheader{2}{pig}
\linerow{faulty & \vb{(!((fileStatusArr = fs.listStatus(path)) == null || fs.isFile(path)))}}
\row{\multirow{2}{*}{manual} & \vb{(!((fileStatusArr = fs.listStatus(path)) == null || fs.isFile(path) ||}} 
\rowline{& \vb{fileStatusArr.length == 0))}}
\row{\multirow{2}{*}{infinitel} & \vb{\inblue{(!(((fileStatusArr = fs.listStatus(path)) == null) || (fs.isFile(path))))}}}
\rowlinespace{& \vb{\&\&((0)<(fileStatusArr.length))}}

\fullheader{2}{tika}
\linerow{faulty & \vb{getContentLength() < getBlockLength()}}
\row{\multirow{2}{*}{manual} & adding variable \vb{continueLoop}:}
\rowline{& \vb{continueLoop \&\& getContentLength() < getBlockLength()}}
\row{\multirow{3}{*}{infinitel} & \vb{\inblue{((getContentLength()) < (getBlockLength()))}}}
\row{& \vb{\&\&((!((this.chmSection.getData().length)==(this.state.getWindowSize())))}}
\rowlinespace{& \vb{||(this.state.getMainTreeTable()!=null))}}

\fullheader{2}{fop}
\linerow{\multirow{2}{*}{faulty} & \vb{(scale < 1 \&\& nextStepFontSize > baseFontSize ||}}
\rowline{& \vb{scale > 1 \&\& nextStepFontSize < baseFontSize)}}
\rowline{manual & adding a \vb{break} statement}
\row{\multirow{4}{*}{infinitel} & \vb{\inblue{(((scale < 1) \&\& (nextStepFontSize > baseFontSize)) ||}}}
\row{& \vb{\inblue{((scale > 1) \&\& (nextStepFontSize < baseFontSize)))}}}
\row{& \vb{\&\&(((FontSizePropertyMaker.FONT\_SIZE\_GROWTH\_FACTOR)+}}
\rowlinespace{& \vb{((FontSizePropertyMaker.FONT\_SIZE\_GROWTH\_FACTOR)-(nextStepFontSize)))<(-1))}}

\fullheader{2}{pdfbox A}
\linerow{faulty & \vb{(amountRead = rawData.read(buffer, 0, Math.min(mayRead,BUFFER\_SIZE))) != -1}}
\rowline{manual & adding wrapping \vb{if}}
\row{\multirow{3}{*}{infinitel} & \vb{\inblue{((amountRead =}}}
\row{& \vb{\inblue{rawData.read(buffer, 0, Math.min(mayRead,BUFFER\_SIZE))) != -1)}}}
\rowlinespace{& \vb{\&\&(filterIndex)<(amountRead))}}

\fullheader{2}{pdfbox B}
\linerow{\multirow{3}{*}{faulty} & \vb{(amountRead = }}
\row{& \vb{read(data, totalAmountRead, numberOfBytes-totalAmountRead)) != -1}}
\rowline{& \vb{\&\& totalAmountRead < numberOfBytes}}
\rowline{manual & modifying \vb{read()} method}
\row{\multirow{4}{*}{infinitel} & \vb{\inblue{((amountRead =}}}
\row{& \vb{\inblue{read(data, totalAmountRead, (numberOfBytes - totalAmountRead))) != (-1))}}}
\row{& \vb{\inblue{\&\&(totalAmountRead < numberOfBytes))}}}
\rowlinespace{& \vb{\&\&((amountRead)==((\inblue{numberOfBytes - totalAmountRead})))}}

\fullheader{2}{uima}
\linerow{faulty & \vb{offset > 0}}
\rowline{manual & modifying loop body}
\rowline{infinitel & \vb{(indent.length())!=(offset)}}

\et{Patches synthesized by $Infinitel$ for the 7 real bugs of our dataset. The code in blue (on screen or a color-printed version) is an expression that is reused from the original patch condition}{realFixes}

\begin{example}
\centering
\begin{subexample}{.25\linewidth}
\begin{batim}[fontsize=\scriptsize]
\inred{- while (condition) \{}
\indarkgreen{+ while (newCondition) \{}




\end{batim}
\captionexample{}{csvLoop}
\end{subexample}
\begin{subexample}{.25\linewidth}
\begin{batim}[fontsize=\scriptsize]
\indarkgreen{+ boolean flag = true;}
\inred{- while (...) \{}
\indarkgreen{+ while (flag && ...) \{}
    ...
\indarkgreen{+    flag = ...;}
\}
\end{batim}
\captionexample{}{tikaLoop} 
\end{subexample}
\begin{subexample}{.23\linewidth}
\begin{batim}[fontsize=\scriptsize]
while (...) \{
...
\indarkgreen{+  if (...) \{}
\indarkgreen{+    break;}
\indarkgreen{+  \}}
\}
\end{batim}
\captionexample{}{fopLoop}
\end{subexample}
\begin{subexample}{.23\linewidth}
\begin{batim}[fontsize=\scriptsize]
\indarkgreen{+  if (...) \{}
    while (...) \{
      ...
    \}
\indarkgreen{+  \}}

\end{batim}
\captionexample{}{pdfboxLoop}
\end{subexample}
\captionexample{Human fixes of: \subrefex{csvLoop} \vb{csv} and \vb{pig}; \subrefex{tikaLoop} \vb{tika}; \subrefex{fopLoop} \vb{fop}; \subrefex{pdfboxLoop} \vb{pdfbox A}}{humanLoopPatches}
\end{example}

\subsubsection{\refitem{adequacy} Synthesis} 

How hard is the code synthesis for each bug? Does the synthesis based on SMT satisfiability scale?
We now concentrate on the code synthesis part based on SMT. 
We look in particular at the number of SMT problems generated.
Recall that there is one SMT problem generated per set of operands to be used for synthesis (row SMT formulations in \reftable{results}).

Let's first look at the three bugs seeded in real code of project \vb{Collections A}, project \vb{Collections B} and \vb{Math}.
The number of SMT problems for \vb{Collections A}, \vb{Collections B} and \vb{Ex.2} is 1. It means the synthesized looping guard directly refer to a boolean variable in the scope (e.g. \qvb{while (notDone)}). Logically, the synthesized code has no operators, which can be seen in the row giving the number of SMT components.
Recall that when the number of SMT components or the number of SMT component types is zero it means that the synthesized condition does not use any operator but only a boolean variable that is present in the scope (e.g. \qvb{while(notDone)}).
For \vb{Math}, the synthesized patch is found for the second SMT problem, using one single operator ($<$). 

We now analyze the code synthesis method for real bugs.
The number of SMT formulations ranges from 4 (for \vb{fop}) to 2 (for \vb{csv}, \vb{pdfbox A} and \vb{uima}).
Those numbers directly refer to the complexity of the synthesized patch, where  several operators are needed, as shown in \reftable{realFixes}.
For instance, the number of SMT formulations of fop is 4 and the corresponding patch contains 5 boolean clauses (fourth row).
In \reftable{realFixes}, note that the operators in blue are not handled by SMT, since it comes from the original loop condition and is evaluated as is.
In our approach, for each new SMT problem, new operators (component) are added for synthesis.
Logically, the greater the number of SMT problems, the greater the number of SMT components and component types, as shown in row ``SMT components'' and ``SMT component types''.
Since at least a component is required to find the patch, it means that there is not a single variable that can be used to describe the completion point alone (such as \qvb{set.isEmpty()}).

\subsubsection{\refitem{limit} Angelic Record} 

Does the 1 million iteration limit used for finding the angelic record have type I errors (false positives)? Does it affect the performance of $Infinitel$?
Recall that the angelic record is the minimum number of executions required to break the infinite loop while the hanging test passes. An angelic record of 0 means that the loop must be skipped, i.e. that the loop precondition is not met.
We use a threshold of one million, which means that if we detect a loop that has more than one million iterations in a single execution, we label it as infinite. 
A false positive would be a loop that indeed needs more than one million iterations, and thus would incorrectly be detected as infinite.
Among the real bugs, there was no false positives. The maximum angelic record is 45 (for \vb{fop}) which is far beyond the maximum value. For seeded bugs, the maximum value is 52 (\vb{Math}).
Indeed, finding the angelic record is the least expensive in terms of time, as shown in \reftable{performance}.

\subsubsection{\refitem{technique} State Observation} 

How does our value collection technique qualify the loop state in each bug? 
The goal of our runtime value collection technique (\refsubsubsection{runtimeCollection}) is to collect the variables that correctly  capture the state of the program. In particular, it must contain the variables that are required to synthesize a correct looping guard.
Now, we look at the elements used in the synthesized looping guard to assess the importance of each phase of our runtime value collection technique.

For seeded bugs, we use extraction by queries, extraction by value from reachable variables, subvalues of the original looping guard and getters.
For real bugs, the patches are given in \reftable{realFixes}. We use visible field access in \vb{csv} (\qvb{tkn.type}). We use recycling of the original looping guard in \vb{pig}, \vb{tika}, \vb{fop} and \vb{pdfbox B} (coloured in blue in the table -- on screen or a color-printed version). We use extraction by queries in \vb{pig} (\qvb{fileStatusArr.length}). We use subvalues of the original looping guard in \vb{pdfbox B} (\qvb{numberOfBytes-totalAmountRead} in the right hand side of the equality component). We use extraction by value of reachable variables (such as the static field \vb{FONT\_SIZE\_GROWTH\_FACTOR} in \vb{fop}). We also use getters in \vb{tika} (e.g., \qvb{this.state.getWindowSize()}).

To sum up, all components of our runtime value collection technique are useful, since that they are all present in the found patches.

\subsubsection{\refitem{idempotence} Idempotence} 

Is there any idempotent loop among the real bugs?
In \refsection{loopTheory}, we have made the analytical arguments that some loops may be idempotent; that is, it is possible to add any arbitrary number of iterations beyond a threshold without breaking the correctness of the computation.
We are now interested in the idempotence of the real bugs. 
The experimental setup is as follows: we simply observe whether the hanging test passes after the maximum number of iterations (one million).

Surprisingly, we remark that the infinite loops of \vb{csv}, \vb{fop}, \vb{pdfbox A}, \vb{pdfbox B} and \vb{pig} are all idempotent.
This aspect could be leveraged during the code synthesis phase: it may occur that the looping guard becomes \qt{easier} to synthesize if more iterations are performed. By easier, we mean it could involve less variables or less SMT components. In this case, this would improve the SMT Solving time, which are particularly high for \vb{csv} and \vb{fop}. However, this optimization remains out the scope of this paper and is left for future work.

\rsection{Discussion}{discussion}

Our approach for automatically repairing infinite loops is built on three assumptions.

The first one is that in each test case there is at most one infinite execution. That is, if the infinite execution is interrupted, any subsequent invocation will be finite. We assume that the once-infinite loop is invoked zero or more times with finite invocations, but at most once with an infinite invocation. We use this assumption in \refalg{findThresholds} because we only probe the angelic record for that single infinite execution.

The second assumption is that the hanging tests have a deterministic execution. Let us assume that in a given test case, a loop is executed several times (for instance by calling $n$ times the method containing it) before entering in the infinite non-terminating mode.
If during the execution of a hanging test the \supscript{n}{th} invocation of an infinite loop is an infinite execution of the loop, then the \supscript{n}{th} invocation of that loop is the infinite invocation on every execution of that test. We use this assumption in \refalg{findThresholds}, because we probe the angelic record in a specific execution number.

Non-determinism in passing tests may also impact the synthesis of the looping guard. We only run each invoking test once to collect runtime values; and the synthesized looping guard guarantees to be correct only for the given input-output pair set. However, suppose a passing test produces two different set of input-output pairs within the \supscript{n}{th} loop execution of two different runs. Namely, $(I_{a},O)$ and $(I_{b}, O)$. That is, the original looping guard evaluates to the same boolean value in the \supscript{n}{th} iteration, but with two different states. Because we synthesize the looping guard using only one of these inputs, say $I_{a}$, there are no guarantees that the looping guard would also evaluate to $O$ for input $I_{b}$. Hence, the synthesized looping guard would make the once-passing test fail intermittently. 

The third assumption is the  collected values are comprehensive enough to synthesize the looping guard.
Although it was sufficient to repair the 14 infinite loops in our dataset, it may not always be the case.
One way to mitigate this problem is to use other method calls when amassing variables for runtime value collection. Another way is to improve our runtime value collection technique. For instance, when we analyze a loop in a method of an \it{anonymous class} (this happens in Java), we do not include instance fields of the anonymous class.

\rsection{Related Work}{related}

One seminal automatic repair technique is $GenProg$ \cite{genprog}.
Genprog is generic by design, and may be applicable for repairing infinite loops.
On the contrary, our repair method addresses a specific defect class (infinite loops). However, $GenProg$ can only find a patch if the repair code already exists in the program, whereas we are able to genuinely synthesize a new expression. 

Dallmeier et al. \cite{dallmeier2009generating} have presented Pachika, a fix generation approach via object behavior anomaly detection. This approach identifies the difference between program behaviors by the executions of passing and failing test cases; then fixes are generated by inserting or deleting method calls. Pachika does not fix loop conditions at all as $Infinitel$ does.

Kim et al. \cite{par} proposed Par, a repair approach using fix patterns representing common ways of fixing common bugs in Java. These fix patterns can avoid the nonsensical patches due to the randomness of some mutation operators.  None of the patterns are specific to infinite loops, and the evaluation does not mention any such bug.

Another recent approach is $SemFix$ \cite{semfix}. The $SemFix$ methodology consists of locating a suspicious assignment or conditional, and then executing the test cases with symbolic execution on that statement. 
The constraints resulting from symbolic execution are then used to identify a state-change that enables the test to pass.
Then, code synthesis is also used to synthesize a code change. 
An extension of $SemFix$ for repairing infinite loops can be envisioned as follows: one can unfold the infinite loop before symbolic execution. This would lead to a path explosion, and it is yet unknown whether this would scale for real programs of size comparable with our dataset.

$NoPol$ is our previous work on automatic software repair \cite{nopol}. $NoPol$ also addresses a specific defect class: wrong conditionals or missing preconditions. 
Our technique $Infinitel$ follows an approach that is similar to the one used in $NoPol$, with two differences. Firstly, $NoPol$ uses spectrum-based fault localization, which is not applicable for infinite loops. We use a completely different technique based on instrumentation to detect the actual infinite loop. Secondly, whereas $NoPol$ finds a boolean angelic value,  we search for an integer angelic record (the loop threshold), representing the minimum number of iterations for a hanging test case to complete and pass.

Regarding one of our aforementioned loop properties, the phenomenon of \qt{idempotent loops}, has also been observed by  \cite{ybranch}: \qt{a specific instance of the loop can iterate for  fewer or greater number of iterations without affecting program output}. However, their goal is completely different from repair, they aim to characterize outcome-tolerant branch instances to discover new ways to enhance the processor's performance. 

Burnim and colleagues \cite{Burnim2009} have presented an approach to detect  infinite loops. They use symbolic execution for reasoning and implements it on top of Java. They only address detection and do not repair the infinite loop as we do in this paper. The same argument applies to the work of Ibing and Mai, which only focus on detection, in a static manner \cite{ibing2015fixed}.

Finally, $Jolt$ \cite{jolt} is an approach to repair infinite loops at runtime. $Jolt$ attaches to an application to audit its progress. It records the program state at the start of each loop iteration. If two consecutive loop iterations produce the same state, $Jolt$ reports that the application is in an infinite loop.
The key difference is that $Jolt$ is at runtime, it simply escapes the loop without changing the looping condition. 
On the contrary, our approach is off-line and based on a hanging test and 
we are able to synthesize a new looping guard.

\rsection{Conclusion}{conclusions}

In this paper, we have proposed a novel method to automatically repair infinite loops. To this end, we have developed static and dynamic source code analysis techniques, along with a code synthesis technique based on SMT problems.
Our method detects the location of the infinite loop, collects execution information about its expected behavior and eventually synthesizes a new loop and correct condition. 

For future work, we will explore whether our framework could be applied to fix a different yet related defect class: wrong loop conditions (which do not necessary result in infinite loops, but to incorrect output) and infinite recursion.

\bibliographystyle{abbrv}
\bibliography{references.bib}

\end{document}